# Electrons Running Away from the Sun Provide the EMF of a Giant Discharge Between the Sun and the Earth

**Philipp I. Vysikaylo**
**Moscow Region State University**
*filvys@yandex.ru*

*Abstract* — In connection with the exit of mankind and its production into open space, the problem of the nature of solar wind (SW) is acute. We have proved that a nonequilibrium inhomogeneous giant gas discharge with huge values of the E/N parameter, which determines the electron temperature, is realized in the heliosphere. This quasi-stationary discharge determines the main parameters of the slow SW in the heliosphere and is the initial background and energy reservoir for all more powerful electrical phenomena in the heliosphere, ionosphere and even in the upper atmosphere of the Earth and the positively charged Sun, connected with the entire heliosphere by reverse electron flows that are unable to leave the positively charged Sun and heliosphere. Our article is devoted to a comparison of the experimental profiles of the global electric field obtained using two methods: 1) by the electron velocity distribution function (in experiments with the **Parker solar probe**) and 2) by the types of positive ions in the SW (according to experiments on the **Prognoz-7 satellite**). We have proved that the Pannekoek-Rosseland-Eddington model does not take into account the important role of high-energy runaway (escape from the Sun) electrons and, accordingly, the duality of electron flows in the heliosphere (from the Sun and towards the Sun). In our model, the slight difference between the opposite currents of high-energy (running away from the Sun) electrons and low-energy (returning to the Sun) electrons is compensated by the current of positive ions and protons from the positively charged Sun – SW (positive ions carry electrons with them).

## I. INTRODUCTION

The study of the interference of gravitational and electrical potentials at the Earth's surface began with the work of William Hilbert [1]. In Hilbert's experiments, the negligible charge of the petals of the same name, constantly received from the thermal plasma of the candle, led to their constant levitation in the gravitational field of the Earth due to Coulomb forces. Such polarization effects - the interference of Coulomb and gravitational forces - have not been studied well enough even in the heliosphere. Our article is devoted to the problem of studying the main parameters of the SW in the inner heliosphere and their coordination with the variety of observed heavy positively charged ions in the solar wind.

The idea of the need to take into account the processes of plasma polarization in stars and star spheres was first formulated by Pannekoek in 1922 [2] and Rosseland in 1924 [3]. They took into account only the difference in gravity in the atmosphere of the Sun due to the difference in masses of electrons and positive ions. Due to this, heavier ions are held by gravity more efficiently than electrons around any star, including the Sun. According to this model, an electric field is created at sizes larger than the Debye radius, compensating the tendency to particle separation in any gravitating plasmoid. The elementary theory of this charge separation for the static heliosphere was created quite a long time ago and it is constantly being modified until now. The corresponding electric field that returns electrons to the star is called the Pannekoek-Rosseland field. In 1926, Eddington analytically estimated the effective positive charge of the Sun ≈ 300 C [4] using this method. Having received negligible electric fields on the surface of the Sun itself, Eddington said, "the effect is absurdly weak" on the surface of the Sun. And most of the astrophysicists believe still that the role of the effective charge of the Sun and galaxies, and, consequently, eddy currents on the charged structures of the Cosmos, is negligible everywhere and is important only on the size of the Debye radius. But, already the Sun's charge of 300 C corresponds to the appearance of reflecting Coulomb mirrors not only for protons, but also for alpha particles. This was first noted in [5].

Polarization models that take into account the global separation of charges due to polarization plasma (in the heliosphere) caused by the difference in the gravitational masses of protons and electrons (main SW components) have been studied in a large number of works with and without taking into account the magnetic field (Lemaire & Scherer 1971; Lemaire & Scherer 1973; Pierrard & Lemaire 1996, 1998; Maksimovic et al. 1997; Scudder 1992; Pierrard, Issautier, Meyer-Vernet and Lemaire 2000; Scudder 2019; Halekas, Whittlesey, Larson, McGinnis, Maksimovic, Berthomier, Kasper, Case, Korreck, Stevens, Klein, Bale, MacDowall, Pulupa, Malaspina, Goetz, and Harvey 2021; and etc.) Detailed references to these scientific literature sources can be found in [6]. However, until now it has not been to explain all the SW parameters and effects from the surface of the Sun to the surface of the Earth, observed in the plasma of the heliosphere, in the ionosphere and in the Earth's upper atmosphere. The fact that many parameters of the SW and the solar surface are related to each other (on the basis of experiments) was confirmed before our work. «It is well known that the time variations of the diverse solar-wind variables at 1 AU (e.g. solar-wind speed, density, proton temperature, electron temperature, magnetic-field strength, specific entropy, heavy-ion charge-state densities, electron intensity etc.) are highly intercorrelated with each other» [7, 8]. According to [8]: «We perform a statistical study with thousands of hours of magnetospheric multiscale observations in the solar wind, comparing the prediction accuracy of the multispacecraft monitor to all of the OMNIWeb single-spacecraft monitors». But, these connections of the observed SW parameters were not properly substantiated theoretically. Therefore, we have proposed a new SW model operating in the entire inner heliosphere from the surface of the Sun to the Earth's orbit [9].

Even the presence of global electric field of Pannekoek-Rosseland should lead to global electric currents in the heliosphere. We prove that global currents flow in the plasma of the heliosphere from the positively charged Sun to the negatively charged Earth, and all the laws of conventional gas-discharge plasma discovered and studied by Stoletov, Townsend, Pashen, ... work in the entire heliosphere to the surface of the Earth.

As we prove, it is more effective in determining the magnitude of the electric field in the heliosphere and the effective charge of the Sun is the change in the types of heavy ions in the SW. This is due to the fact that in the region of stars in electric fields in the presence of a gravitational potential, the direction of motion of positive ions is discriminated against by the parameter $\zeta = Z/M$ [5]. Here $Z$ is the charge number of the positive ion, M is its mass number. For the Sun $\zeta = 0.107$ [9]. At $\zeta < 0.107$ positive ions fly towards the Sun, and at $\zeta > 0.107$ they fly away from the Sun. At present, we know two methods for determining the electric field profile in the heliosphere: 1) from the types of heavy ions in the SW [5, 9] and 2) from the change in the velocity distribution function of electrons [6]. The purpose of this work is to compare the results of such experimental studies of electric fields in the heliosphere. The verification of the obtained electric field strength profiles by these various methods will be carried out on the basis of an explanation of the available experimental data on the entire set of phenomena in the heliosphere, the solar corona and the upper layers of the Earth's atmosphere.

## II. GENERAL AND DIFFERENCES IN OUR AND PANNEKOEK-ROSSELAND-EDDINGTON'S MODELS

As noted earlier, Pannekoek 1922 put forward the idea that the difference in masses of protons and electrons should lead to charge separation and the generation of polarizing electric fields in the heliosphere. In order to maintain quasi-neutrality in the presence of mass-dependent gravitational forces, electric fields must exist (Pannekoek 1922). Given the large thermal velocity of the light (easy) electrons, a considerable electric potential drop should exist between the solar corona and 1 AU (Lemaire & Scherer 1971). A class of "exospheric" models (Lemaire & Scherer 1973) posits that this electric field accelerates the solar wind ions from the corona, with the pervasive non-thermal nature of the electron distribution increasing the efficiency of this acceleration (Pierrard & Lemaire 1996; Maksimovic et al. 1997; Scudder 1992). This electric field may self-consistently generate the nonthermal features of the electron distribution through a runaway process (Scudder 2019). The main relations in these works are the following equations [6]:

$\Sigma Z_i n_i = n_e$ (1)
$\Sigma Z_i n_i v_i = n_e v_e$ (2)

$Z_i, n_i, v_i$ - charge, concentration and velocity of positive ions; $n_e$, $v_e$ – concentration and velocity of electrons.

The electric field that returns electrons to the star according to the model (1)-(2) is called the Pannekoek-Rosseland field. The movements of charged particles according to Model 1-2 are shown in Fig.1a.

Model (1) - (2) does not take into account the duality of the electron current (Fig.1a). In this model, all streams of charged particles come from the Sun and there are no streams affecting the Sun and its charge. The cumulation of electrons to the Sun in the Pannekoek-Rosseland electric field is not taken into account in their model. This is an error.

Our model takes into account duality of the electron current (Fig.1b). High-energy electrons can even leave the potential well - the positively charged Sun and the heliosphere, while the less energetic ones are slowed down by the electric field in the heliosphere and return back to the positively charged Sun.

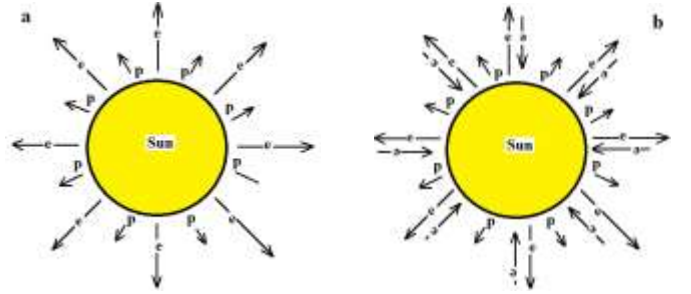

**Fig.1.** Schemes of the movement of electrons - e and positively charged ions - in the heliosphere: **a)** – according to model (1-2) and **b)** – according to our model (3-4) (*p* - positively charged ions with $Z/M \geq 0.107$).

In our model, quasi-constant fluxes of electrons to the positively charged Sun are practically equal to the flux of high-energy electrons from the Sun and differ by the value of the current of positive ions from the Sun. In contrast to model (1) - (2), we have:

$n_{e2} v_{e2} = n_{e1} v_{e1} + \Sigma Z_i n_i v_i$ (3)

$n_{e2}$, $v_{e2}$ - concentration and speed of high-energy electrons escaping from the heliosphere; $n_{e1}$, $v_{e1}$ - concentration and speed of electrons concentration and speed of electrons returning to the Sun.

Instead of the identical equality (1), we believe that there is a weak violation of electroneutrality on the Sun and throughout the inner heliosphere. The presence of free electrons in the Sun and heliosphere and their high mobility lead to the escape of some high-energy electrons from the Sun. The higher the temperature of the electrons, the greater the part of the electrons capable of leaving the heliosphere (Fig.2). This leads to an increase in the effective charge of the Sun and a corresponding increase the velocities $v_{e1}$, $v_{e2}$, $v_i$ and penetration of the electric field of the effective charge of the Sun into the entire heliosphere, where the electron temperature is significantly higher than on the Sun's surface and reaches 1-2 million degrees in the Sun's corona. In Fig.1, we took into account that the main current of positive ions is associated with protons and alpha particles. (Positive ion fluxes are also divided by the interference of gravitational and Coulomb potentials into fluxes to the Sun at Z/M < 0.107 and fluxes from the Sun at Z/M ≥ 0.107, see Fig.3 and section III). At this temperature, the formation of a compensatory layer above the positive charge of the Sun does not occur, and the electric field of this quasi-constant charge of the Sun penetrates into the entire heliosphere. This field heats electrons (this leads to the growth of $v_{e1}$ and $v_{e2}$ in (3)) and accelerates positively charged ions, protons and alpha particles from the Sun (at $Z/M \geq 0.107$). The plasma wind of positive ions (mainly protons) carries electrons with it. This is how the solar wind from the Sun is formed according to our model. This model takes into account not only the flow of energy, masses and impulses from the Sun, but also flows to the Sun. The dynamic cumulation of returning electrons ensures the connection of the entire positively charged heliosphere with the parameters of the Sun's surface and its effective charge and the charge of the entire heliosphere and even planets.

Consider the condition of quasi-neutrality and penetration of electric fields in a plasma with a global current. In the scientific and educational literature, various and contradictory definitions of plasma quasi-neutrality are given. The concept of "quasi" is almost. So, quasi-permanent – almost permanent. Without the prefix "quasi", the plasma becomes electrically neutral or, according to the Poisson equation, the density of the space charge in the plasma must be zero everywhere. Therefore, it is necessary to apply equation (1) carefully when modeling a discharge in the heliosphere. In the general case, a reasonable definition of the quasi-neutrality of a simple plasma is:

$$\alpha_i = (n_i - n_e)/N \ll 1 \quad (4)$$

Where $n_e$ is the electron concentration; $n_i$ is the concentration of positive ions; $N$ is the nucleon concentration (protons and neutrons in atoms, molecules, etc.) The definition of the electrical neutrality violation parameter (4) follows from the works of Eddington [4] and Shklovsky, see details in [5]. This parameter determines the ratio of Coulomb forces to gravitational ones. The question remains open: how many times is $\alpha_i$ less than 1 in the entire heliosphere and in the Sun?

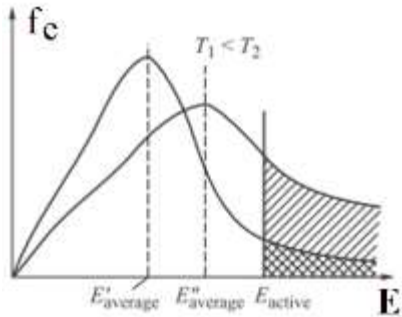

**Fig.2.** Electron energy distribution functions depending on temperature

### III. CALCULATIONS OF THE EFFECTIVE POSITIVELY SUN CHARGE AND POSITIVE ION PARAMETERS IN THE SOLAR WIND

In the heliosphere, two forces act on positively charged ions: the force of gravity and the force of the electric field. If we assume that the electric field is determined by the uncompensated charge of the Sun, then we can estimate the effective charge of the Sun from the ionic composition of positive ions in SW [9]. Experimenters in [10, 11] found that the following positive ions are observed in SW: C4+, O5+, Ne8+, Mg6+, Si6+, Fe6+, …, Fe13+… ($Z/M$ = … = 0.33; 0.31; 0.39; 0.24; 0.21; 0.107…). In SW there are heavy positive ions ionized 4 or more times. The absence of positive Fe5 + ions in SW (Fig.3) gives us the opportunity to assume that for them the forces of Coulomb repulsion from the positively charged Sun are less than the force of gravitational attraction. Hence the effective positive charge of the Sun is $Q_S = 4\pi\varepsilon_0 \cdot G \cdot M_S \cdot M_{Fe+6}/(6e) \approx 1,400$ C [9]. $M_{Fe6+}$ – ion mass of Fe6+. With this effective charge of the Sun, all positive ions with $Z/M \geq 0.107$ fly from it, since for them the gravitational forces are less than the Coulomb forces [9].

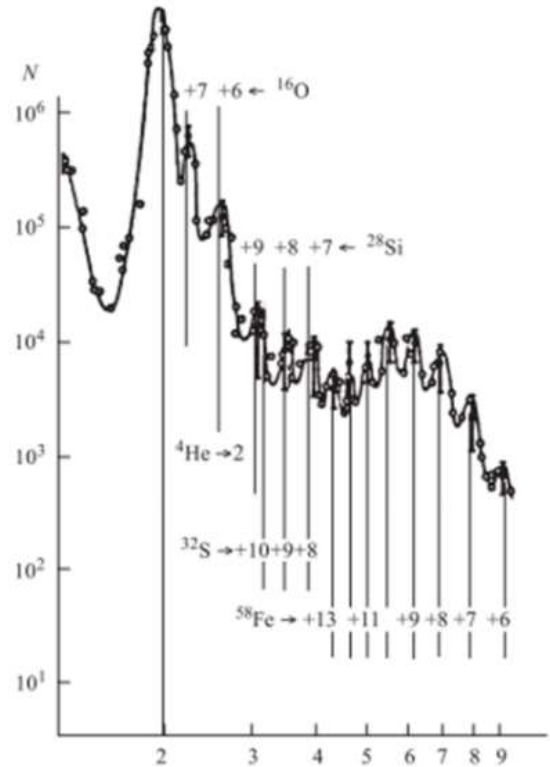

**Fig.3.** Typical ion spectrum in the solar wind, measured on the Prognos (Forecast)-7 satellite in 1978-1979. On the horizontal axis – the ratio of the mass of particles m to its charge in 10 s. The numbers with the sign "+" denote the charge number of the ion. The proton peak with $Z/M = 1$ is not shown, since it exceeds the peak of α–particles by more than an order of magnitude [11].

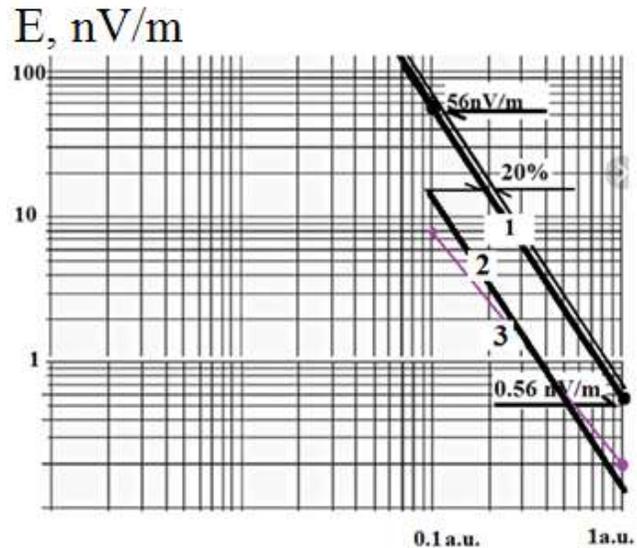

**Fig. 4.** Calculations of electric fields in the heliosphere: 1- according to the spectrum of positive ions in the solar wind [9] ($Q_S$ = 1,420 C; error less than 20%); 2 - according to the Eddington's calculations [4]; 3 - according to the calculations of the electric field based on experimental observations of the electron distribution function using the Parker solar probe [6] ($Q_S$ = 200 ÷ 500 C, the dependency 3 is built by J.S. Halekas).

If there are no negative charges in the heliosphere up to the Earth that could implement Debye's shielding of the effective charge of the Sun, then knowledge of the parameters of the Sun allows us to determine: 1) the electric field strength on the solar surface at $E(R_S) \approx 2.7 \cdot 10^{-5}$ V/m; 2) the electric field profile in the entire heliosphere $E(R) = KQ_S/R^2$ (Fig.4).

The results obtained by us when taking into account the types of ions in SW in [9] are compared with the results obtained by Eddington [4] (electric field strength and the solar charge differ by 4.2 times) and J. S. Halekas et al. [6] in Fig.4, depend 3 (the intensity of the electric field and the charge of the Sun differ from 7 in the region of the Sun's corona up to 2.8 times and in the region of the Earth's rotation).

## IV. THE ELECTRON TEMPERATURE PROFILE IN THE HELIOSPHERE

According to Paschen's law, the energy of a charged particle is determined not only by the $E(R)$ – electric field (EF) strength, but also by the path length, where a charged particle gains energy in the EF without collisions, i.e., by the $N(R)$ particle number density or by the $E/N$ parameter. This law is used here to describe "mysterious" phenomena in heliosphere inhomogeneous in the $N(R)$ [12-15]. If an increase in $E$ leads to an increase in the temperature of electrons, then a decrease in the density of the number of particles - $N$ (participating in the scattering of electrons and ions energy) leads to the same effect.

As we can see from the Table, the data in various literary sources on the density of the number of particles (N1 or N2), on which the relaxation of the energy of charged particles in an EF occurs, are in good agreement in the central region of the inner heliosphere. A significant difference between the N2 data in [12] and the N1 data in [13] is observed in the near corona of the Sun ($R \sim 1000$ km) and in the region of 1 AU. Here the difference reaches $10^2$.

We construct the electron temperature profile – $T_e(R)$ in the entire heliosphere based on the Nernst — Townsend (Einstein-Smolukhovsky) relation [9]:

$$T_e = e\, D_e/\mu_e \sim (E/N)^\zeta \qquad (5)$$

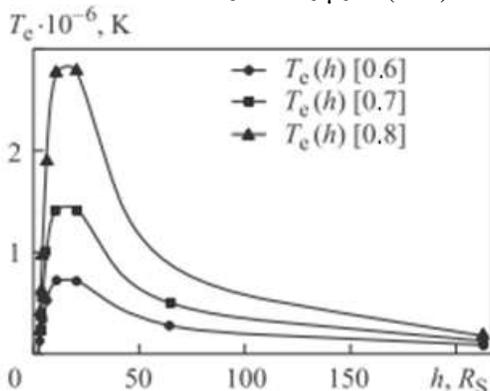

**Fig.5.** Electron temperature dependence on distance to the Sun at the [ζ] parameter values with the Sun charge of 1,400 C; $h > 10\,000$ km [9].

The calculations by the formula (5) for the data in the Table and Fig.5 were carried out under the assumption that the concentrations of electrons are equal to the concentration of hydrogen molecules in the entire heliosphere. This assumption is reasonable in the region of the heliosphere at distances greater than $20 \cdot R_S$ and to the Earth's orbit. At distances closer to the Sun, $R < 20 \cdot R_S$, the dissociation of hydrogen molecules and an increase in the proton concentration should be taken into account. If we take into account that in the region of the surface and the corona of the Sun, the neutral gas is atomic hydrogen or even proton gas, then the temperature profiles will rise even more sharply in Fig.5 near the Sun. Such an allowance will lead to a decrease in the breakdown electric fields to E/N on the order of 0.06 Td and below up to 1000 km from the Solar surface (see Table). This problem will be considered in more detail in future publications.

**Table**
**Dependencies of profiles of particle number densities in the heliosphere from different sources $N_1$ [13], $N_2$ [12,14,15] and E/N [Td] (set by the solar charge) depending on the altitude above the Sun surface**

| R, km; /R_S | $N_1$ [11], cm$^{-3}$ | $N_2$ [10], cm$^{-3}$ | E/N$_1$, Td | E/N$_2$, Td | $T_e$, K/ζ |
|---|---|---|---|---|---|
| 0 | | $1.5 \cdot 10^{17}$ | | | 6 430 |
| 91 | | $10^{17}$ | | | 5 410 |
| 200 | | | | | 4 170 [12] |
| 320 | | $1.3 \cdot 10^{16}$ | | | 4 560 |
| 520 | | $2.4 \cdot 10^{15}$ | | | 4 190 |
| 560 | | $1.6 \cdot 10^{15}$ | | | 4 180 |
| 0/1 | $3.9 \cdot 10^{15}$ | | $6.7 \cdot 10^{-6}$ | | |
| 1,000 | $3.1 \cdot 10^{13}$ | | $8.4 \cdot 10^{-5}$ | | |
| 1,004 | | $2.5 \cdot 10^{13}$ | | $1.06 \cdot 10^{-5}$ | 5 750 |
| 1,580 | | $4 \cdot 10^{11}$ | | 0.066* | 7 150 |
| 1,900 | | $1.3 \cdot 10^{11}$ | | 0.2* | 375* |
| 2,0 | $6.3 \cdot 10^{12}$ | | $4.2 \cdot 10^{-3}$ | | |
| 2,0 | | $10^{11}$ | | 0.26* | 460* |
| 2,1/1.003 | | $10^9$ | | 26.6* | 11,553* |
| 3,000 | $2.0 \cdot 10^{12}$ | | $1.3 \cdot 10^{-3}$ | | 30*/0.7 |
| 3,500/1.005 | | $6 \cdot 10^8$ | | 44* | 16,697* |
| 4,000 | $7.9 \cdot 10^{11}$ | | $3.3 \cdot 10^{-3}$ | | |
| 6,000 | $2.51 \cdot 10^9$ | | 10.6 | | 6,164*/0.7 |
| /1.01 | | $3.2 \cdot 10^8$ | | 83* | 26,035*/0.7 |
| 7,000 | | $5 \cdot 10^9$ [12] | | | 25,000* |
| 8,000 | $10^9$ | | 26 | | 11,5*/0.7 |
| 10,000 | $6.3 \cdot 10^8$ | | 40.8 | | 15,84*/0.7 |
| 14,000 | | $5 \cdot 10^8$ [12] | | | 300,000* |
| 15,000 | $2.0 \cdot 10^8$ | | 126 | | 34,872*/0.7 |
| /1.03 | | $2 \cdot 10^8$ | | 133* | |
| /1.06 | | $1.6 \cdot 10^8$ | | 166* | |
| 70,000 | $7.9 \cdot 10^7$ | | 279 | | 60,833*/0.7 |
| 70,000 | | | | | $2 \cdot 10^6$ [13] |
| /1.1 | | $10^8$ | | 233* | |
| /1.2 | | $6.7 \cdot 10^7$ | | | |
| /1.2 | | $4 \cdot 10^7$ [12] | | | $1.510^6$*[12] |
| 280,000 | $1.2 \cdot 10^7$ | | 1,079 | | 156,794*/0.7 |
| /1.3 | | $2 \cdot 10^7$ | | | |
| 420,000 | $5.0 \cdot 10^6$ | | 2,074 | | 247,732*/0.7 |

| | | | | |
|---|---|---|---|---|
| /1.4 | | $8 \cdot 10^6$ | | |
| 700,000 | $1.6 \cdot 10^6$ | | 4,204 | 406,236*/0.7 |
| /2 | | $3.3 \cdot 10^6$ | | |
| /2 | | $10^6$ [13] | | |
| /2.5 | | $10^6$ | | |
| 1 400,000 | $4.0 \cdot 10^5$ | | 7,421 | $0.9 \cdot 10^{6*}/0.7$ |
| /3 | | $2.8 \cdot 10^5$ | | |
| /4 | | $1.5 \cdot 10^5$ | | |
| /5 | | $10^5$ [13] | | |
| 2 800,000 | $6.3 \cdot 10^4$ | | 16,851 | $10^{6*}/0.7$ |
| /5 | | $5.0 \cdot 10^4$ | | |
| /6.6 [12] | $3.1 \cdot 10^4$ | ?? | | |
| 6 200,000 | $10^4$ | | 26,45 | $1.5 \cdot 10^{6*}/0.7$ |
| /10 | | $10^4$ | 25,7 | |
| /10 | | $10^4$ [12] | | |
| /11 | | $10^4$ [13] | | |
| 13 000,000 | $2.5 \cdot 10^3$ | | 26,450 | $1.5 \cdot 10^{6*}/0.7$ |
| /20 | | $1.6 \cdot 10^3$ | 41,260* | |
| /50 | | $1.6 \cdot 10^2$ | 66,540* | $2.8 \cdot 10^{6*}/0.7$ |
| /50 | | $10^2$ [12] | | |
| 44 000,/65 | $10^3$ | | 6,290 | $0.55 \cdot 10^{6*}/0.7$ |
| /100 | | $3.1 \cdot 10^1$ | 85,830* | |
| 150 000,000 | $6.3 \cdot 10^2$ | | 909 | $0.18 \cdot 10^{6*}/0.7$ |
| /215 | | 5 | 114,500* | $4 \cdot 10^{6*}/0.7$ |
| /215 | | 2.5 [12] | | $6.4 \cdot 10^{6*}/0.7$ |
| /214 | | | | $10^5$ [13] |

## V. CONCLUSIONS

As a result of experiments carried out with the help of the «Parker» solar probe (2018-2022), the question of the presence of electric fields in the heliosphere was definitely resolved in the direction of their presence. Using data on the grades of positive iron ions obtained from the «Forecast-7» satellite (1978-1979), we calculated in [9] the electric field profile in the entire heliosphere, significantly exceeding the values obtained in [6] (Fig.4).

Heavy ions are formed in the corona, and their state does not change when moving in SW, due to the absence of collisions in the heliosphere. Consequently, SW ions carry information about conditions in the solar corona. SW observations are of practical importance. According to the types of observed positive ions in SW, according to [9], we can determine the magnitude of the effective quasi-constant positive charge of the Sun. Knowing the profile of the electric field $E(R)$, we can calculate the energy of charged particles born in the sun's corona and compare it with the energy of particles observed experimentally in the Earth's orbit (in the slow solar wind). These comparisons are in good agreement with an effective charge of the Sun of 1,400 C [9]. At the Sun charge determined in [4] (Fig.4, line 2), the proton energy is 4.5 times less than that observed in a slow SW, and when determined in [6] (Fig.4, line 3), it is 7 times less than that observed in a slow SW [17]. Verification of mathematical models of the heliosphere should be carried out in a comprehensive manner, taking into account all the facts observed by us experimentally, taking into account the types of positive ions observed in SW (Fig. 3). In all models (see ref. in [6]), the types of positive ions were not taken into account in determining the effective Sun charge and the incorrect model 1-2 (Fig.1*a*) was used, not 3-4 (Fig.1*b*) as in our case.

For the first time as a result of analytical studies of toroidal current in the heliosphere global electric circuit, according to model 3-4, we have calculated the parameters of the heliosphere and verified by experimental observations:

1) effective positive Sun charge is 1,400 C;
2) with such a positive charge, the Sun is able to reflect all positively charged particles with the ratio (of the charge number – $Z$ to the mass number – $M$) $Z/M \geq 0.107$. These are such positively charged particles as: protons, alpha particles and multiply ionized ions of heavy atoms (C4+, O5+, Ne8+, Mg6+, Si6+, Fe6+, …, Fe13+ generated in the solar corona), observed in the SW [10,11,17];
3) protons and alpha particles velocities arising at distances from the Sun of $(10 \div 30) \cdot R_S$. They are accelerating in EF of the positively charged Sun up to 400 km/s and more in the Earth area;
4) non-compensation coefficient of the Sun positive charge $\alpha_i \approx 7.5 \cdot 10^{-36}$, i.e., for $10^{36}$ compensated nucleons there are only 7.5 electrons that left the Sun far beyond the problem scope (for example, the Solar system);
5) $E/N$ parameter profile for heliosphere, which determines the breakdown conditions and the electron temperature, the $E(R)$ EF strength profile was calculated according to the Coulomb law, and the profile of nucleon density in heliosphere was taken from [12-15];
6) the $T_e$ profile in the entire heliosphere (from the Sun to the Earth) was calculated using the Nernst — Townsend (Einstein — Smoluchowski) relation for the E/N profile according to (5) [9]. The $T_e$ profile (Fig.5) with $\zeta = 0.7$ is in good agreement with experimental observations, see Table.

The electric field of the positively charged Sun penetrates into the entire heliosphere, into the ionosphere and even into the Earth's upper atmosphere, since constant streams of high-energy electrons are formed in the entire heliosphere, as well as in the Sun itself, escaping from the Sun and the hot heliosphere, and they are not enough to form the Debye compensation layer of the positive charge of the Sun.

In the Sun and throughout the heliosphere, when exchanging energy of electrons in electron-electron collisions, a stream of highly energetic electrons constantly escaping from the Sun and the heliosphere is formed. This process plays the role of EMF in the heliosphere between the positively charged Sun and the Earth negatively charged up to 500kC. The negative charge of the Earth is due to the capture of high energy electrons by the Earth.

For a laboratory gas discharge, the particle number density usually changes slightly, and in the heliosphere up to $10^{17}$ times from the surface of the Sun to the Earth's orbit. For the heliosphere, where the density of gas particles, $N$, falls faster than the electric field strength from the distance to the Sun, a solar charge of 1,400 C leads to the generation of breakdown values of $E/N$ already at altitudes of the order of 1000 km. At altitudes of 10-30 solar radii, a non-equilibrium gas-discharge plasma is formed with huge values of the $E/N$ parameter and, consequently, electron temperatures up to 1 - 2 million degrees (see Table, Fig.5). At such temperatures, global toroidal currents are formed in the entire heliosphere (Fig.1*b*), caused by the escape of high-energy electrons from the Sun [9]. As proved in our works, this is due to a sharp drop in the number of particles on which the relaxation of the electron energy - $N$ occurs with increasing $R$.

As we prove, in the inhomogeneous heliosphere several types of discharges are simultaneously co-organized into a single discharge with different characteristic Debye radii from $R_D \sim 10^{-8}$ m (in the region of the Sun) to $R_D \sim 10$ m in the region of the Earth's orbit, with electron temperatures from $T_e$ = 4,170K (at the surface of the Sun) to $T_e \sim$ 1-2 million K at distances of (20-30)·$R_S$ and $T_e \sim 10^5$K in the region of the Earth's orbit. It is usually believed that a quasi-neutral plasma (with a small Debye screening radius) should lead to a noticeable deviation of the electric field in the heliosphere from the Coulomb one. However, an increase in the temperature of electrons with distance from the Sun leads to the impossibility of forming a compensating layer in the heliosphere due to the departure of high-energy electrons from the heliosphere. As can be seen from experimental studies [6] (Fig. 4, dependence 3), the effective charge of the Sun increases with distance from the Sun in the heliosphere by 2.5 times. Electron flows from the Sun in experiments [6] are closed from recording devices. Possibly, this leads to a significant underestimation of the effective electric fields. It would be necessary to simultaneously experimentally measure the types of positive ion fluxes in the entire heliosphere in accordance with the methods [10,11,17]. The types of positive ions established in the SW make it possible to determine the effective charge of the Sun or its intensity profile according to the method [9].

Our model (Fig. 5, Table) assumed the presence of non-ionized hydrogen molecules in the entire heliosphere. This made it possible (according to available laboratory studies) to use the most successful approximation of the dependence of $T_e$ on E/N in equation – (5), in the range of 1-200 Td and at large values of the parameter E/N (in the heliosphere).

According to our ideas, a giant discharge is realized in the heliosphere between the positively charged Sun and negatively charged planets (for example, the Earth). In our model, the Sun is an analogue of the anode, and the negatively charged Earth is the cathode. In the heliosphere, the negatively charged Earth (as a cathode) is surrounded by the positively charged ionosphere of the Earth - an analogue of a positively charged cathode spot, on which there is a significant drop in the external potential - cathode drop. The large charge of the Earth (500 kC [16]) in comparison with the effective charge of the Sun (1,400 C [9]) can be explained by a significant voltage drop across the positively charged ionosphere. The positively charged heliosphere itself is analogous to the positive column of plasma in a laboratory discharge. For this reason, the heliosphere is positively charged and does not have a compensation layer for the positive charge of the Sun at distances from the Sun at the size of the Debye radius. In a laboratory gas discharge, the EMF is usually outside the discharge gap, and in the heliosphere, the EMF of such a giant discharge is due to the escape of high-energy electrons from the entire heliosphere and from the Sun and the processes of electron-electron collisions, leading to the transfer of energy to the tail of the electron energy distribution function (in the region of energies characteristic of runaway electrons [9]). The intrusion of high-energy electrons escaping from the Sun into the planets leads to a negative charge of the planets (in particular, the Earth up to 500 kC [16]). If we take into account the presence of such a global gas inhomogeneous nonequilibrium discharge (Fig.1b), one can explain many phenomena observed in the heliosphere, in the ionosphere and in the Earth's upper atmosphere, which are similar to those in a laboratory discharge [18]. Usually, electrical phenomena in the atmosphere are associated with the water cycle in the Earth's atmosphere [16]. Now we can analyze phenomena such as blue jets, jets and sprites and develop their models based on our global discharge model in the heliosphere between the Sun and the Earth. Note that these phenomena are realized in the upper atmosphere at altitudes of about 40 km, where there is practically no water vapor [16].

The role of high-energy electrons escaping from plasma structures in the formation (self-cumulation due to the generation of dynamic surface tension forces) and propagation of plasma structures directed at the anode has already been well studied in the scientific literature [19]. In this case, the electrons are accelerated by the electric field and fly out of the plasma structure against the direction of the electric field – $E$, without collisions leading to the loss of their energy. The escaping high energy electrons provide preionization and transition of the non-activated gas medium into plasma.

In contrast to [19], in our works it was proved for the first time that a sharp decrease in the concentration of particles - N, on which the energy of electrons (escaping and returning to the heliosphere) is dissipated, is a very important feature of the global discharge in the heliosphere (Fig.1b). We see that at distances from 1000 km from the Sun and up to 10-30 radii of the Sun (Fig.5 and Table), there is a sharp increase in the E/N parameter. This is due to a sharp decrease in the concentration of particles – N (inhomogeneous due to the presence of gravitational forces). This phenomenon (Fig.5, Table), observed in the heliosphere, proves the asymmetry of the influence of the electric field – E and density – N on the plasma parameters in the heliosphere [9,18], as in the laboratory gas discharge (see the works of Stoletov, Pashen, Townsend). The discoveries of electric field shock waves (an increase in electric field strength and volume charge density due to a weak violation of electroneutrality) in semiconductors (the Gann effect) and their visualization and theoretical description in gas-discharge plasma in [20] allow us to assert that a standing shock wave of E/N (an electric field – $E$, reduced to the N – density of the number of particles on which electron energy relaxation occurs) is formed in the heliosphere. In this region, due to the huge values of E/N, there is a huge release of heat and heating of all components of the heliospheric plasma. This is the standing shock wave of the reduced electric field - E to the density of the number of particles – N. This phenomenon requires further studies.

Magnetic phenomena were reported in China more than 2 thousand years ago. Electrical phenomena have been explored since Faraday's experiments, only 150 years ago. In a gas discharge, the importance of the parameter E/N (measured in townsends) has long been discovered. In astrophysics, magnetic phenomena are still studied, not electrical ones. In our spherically symmetric model, magnetic forces are equal to zero in the heliosphere and arise only in the region of planets that entrain charged SW particles, thereby forming circular currents. This is a separate task.

We have two methods for experimental determination of the local electric field (Fig. 3, dependences 1 [9] and 3 [6]). Based on these methods, we can determine the profiles of all parameters of charged particles in the entire undisturbed heliosphere. Based on the experimental results of [6] ($Q_S \sim$ 200-500 C), we cannot explain: 1) the presence of positive ions Fe +6,7,8; 2) the absence of positive Fe+4,5 ions in the SW; 3) electron temperature of 1-2 million degrees in the heliosphere; 4) the speed of charged particles in SW in the region of the Earth's orbit. All parameters require an increase in the effective charge of the Sun by 5-7 times. Our quantitative studies of the nonequilibrium inhomogeneous heliospheric plasma with runaway electrons have shown that the method [6] applied on the «Parker» solar probe requires significant modification and taking into account the contribution of high-energy electrons escaping from the Sun and the heliosphere.

At the same time, the second method [9], developed by us on the basis of determining the types of positive ions in the SW (see [10, 11]), gives the values of the SW parameters that make it possible to explain the entire spectrum of dynamic phenomena known to us in the heliosphere and even the upper layers of the atmosphere. Positive ions gain energy from $10 \cdot R$s to the Earth's orbit in the electric field of the positively charged Sun and heliosphere. In this case, the heliospheric electric field in the region of the Earth's motion is not large, it is about 0.6 nV/m (fig.4).

Thus, we have verified the model of a quasi-constant slow solar wind by various experiments [5,9,18]. This model of an inhomogeneous nonequilibrium plasma is acceptable for a theoretical description of the initial stage (initial background) of an already strong self-organizing solar wind, which manifests itself in the form of pheloments in the solar region and various plasmoids in the form of elves, jets, sprites and other structures in the region from the surface of the Sun to the upper atmosphere of the Earth.

The results of this work were accepted for presentation at the 240th meeting of the American Astronomical Society in Pasadena, California, June 12–16, 2022. It can be found as an i-poster in [21].